IAC-24-E10.1.1

# A Mission to Demonstrate Rapid-Response Flyby Reconnaissance for Planetary Defense

Nancy L. Chabot[a]*, Justin A. Atchison[a], Rylie Bull[a], Andrew S. Rivkin[a], R. Terik Daly[a], Ronald-L. Ballouz[a], Olivier S. Barnouin[a], Andrew F. Cheng[a], Carolyn M. Ernst[a], Angela M. Stickle[a], Evan J. Smith[a], Joseph J. Linden[a], Benjamin F. Villac[a], Jodi R. Berdis[a], Dawn M. Graninger[a], and Sarah Hefter[a]

[a] *Johns Hopkins University Applied Physics Laboratory, Laurel, MD, 20723, USA.*
\* Corresponding Author, nancy.chabot@jhuapl.edu

**Abstract**

International and U.S. strategies and protocols have identified the need to develop rapid-response spacecraft reconnaissance capabilities as a priority to advance planetary defense readiness. A space-based reconnaissance response is recommended for potential impactors as small as 50 m, making these small objects the most likely to trigger a space-based response and the ones that drive the reconnaissance capabilities needed. Even following the successful completion of the NEO Surveyor mission and Rubin Observatory survey efforts, roughly half of the 50-m near-Earth object (NEO) population will remain undiscovered. As a result, 50-m impactors may not be found with long warning times, and a rapid-response flyby mission may be the only reconnaissance possible. To develop a robust flyby reconnaissance capability for planetary defense, four major requirements are defined for a demonstration mission: 1) Enable a flyby of >90% of the potential asteroid threat population; 2) Demonstrate the flyby reconnaissance for a ~50 m NEO; 3) Obtain the information needed to determine if and where it would impact the Earth; and 4) Determine key properties of the asteroid to inform decision makers. To meet these requirements, flyby speeds of up to 25 km/s and approach solar phase angles as high as 90° have to be accommodated within the mission's capabilities, which becomes an even greater challenge when the mission must obtain the needed reconnaissance measurements for a small, 50-m object. However, as commonly noted in the community, in planetary defense, you don't pick the asteroid — the asteroid picks you. Thus, a planetary defense flyby reconnaissance demonstration mission is not about just flying by an asteroid, but rather it is about developing a robust capability for the objects that are most likely to require a short-warning-time, space-based response, and addressing this challenge is necessary to advance our planetary defense preparedness capabilities.

## 1. Introduction

In 2017, the United-Nations-endorsed Space Mission Planning Advisory Group (SMPAG) recommended three threshold criteria that would warrant starting spacecraft mission options planning in response to a potential near-Earth object (NEO) impact threat [1]: 1) Earth impact is predicted to be within <50 years, 2) impact probability is assessed to be >1%, and 3) the object is characterized to be >50 m in size (or absolute magnitude (H) of <26 if only brightness data can be collected). In 2021, the SMPAG criteria were used to inform the United States Report on Near-Earth Object Impact Threat Emergency Protocols (NITEP) [2] to define the benchmark for recommending space-based reconnaissance. NITEP recommends that the U.S. consider executing a spacecraft reconnaissance mission in any scenario that meets the SMPAG criteria and also has sufficient time to conduct the mission prior to the predicted impact. NITEP notes that *"such a mission can provide information critical to determining whether prevention is necessary and, if so, to enabling its success."*

These SMPAG [1] and NITEP [2] thresholds are set by recognizing that even a 50-m object is capable of local devastation with the potential for regional effects, and these potential consequences warrant space-based action. There are estimated to be ~230,000 asteroids that are roughly 50 m in diameter in the NEO population, and a 50-m object impacts the Earth roughly every thousand years, more frequently than larger objects [3]. Thus, a 50-m-diameter object is the smallest for which a reconnaissance spacecraft mission is recommended by international and US protocols, and it is also the most likely case to be encountered to trigger a space-based response.

Additionally, even after the successful completion of NEO Surveyor operations [4] and Rubin Observatory survey efforts [5], roughly half of the 50-m NEO population will still be left undiscovered. As a result, many 50-m impactors may not be found with long warning times, in contrast to objects ≥140-m, for which >90% will be known and tracked following the completion of NEO Surveyor's mission [4]. Consequently, a rapid-response flyby mission may be the only viable option for a reconnaissance mission to a 50-m object discovered with short warning time.

With this motivation, the 2023 U.S. Decadal Survey for Planetary Science and Astrobiology [6] recommended that *"the highest priority planetary defense demonstration mission to follow DART and NEO Surveyor should be a rapid-response, flyby reconnaissance mission targeted to a challenging NEO, representative of the population (~50-to-100 m in diameter) of objects posing the highest probability of a destructive Earth impact."* The 2023 U.S. National Preparedness Strategy and Action Plan for Near-Earth Object Hazards and Planetary Defense [7] has six top-level goals, with Goal 3 being to *Develop Technologies for NEO Reconnaissance, Deflection, and Disruption Missions*. In particular, one of only two short-term actions listed to support Goal 3 is "*3.2 Create plans for the development, testing, and implementation of NEO reconnaissance mission systems.*" The 2023 NASA Planetary Defense Strategy and Action Plan [8] also adopts the same priority actions for Goal 3 as are listed in the U.S. National Preparedness Strategy and Action Plan [7].

Overall, there is strong international recognition of the value provided by spacecraft reconnaissance of a potential NEO impact threat and support for developing the needed NEO reconnaissance capabilities to advance planetary defense readiness. In this work, we address this planetary defense priority by determining the requirements for a planetary defense rapid-response flyby reconnaissance demonstration mission. As commonly noted in the community, in planetary defense, you don't pick the asteroid — the asteroid picks you. Thus, the motivation for a planetary defense flyby reconnaissance demonstration mission is not simply flying by an asteroid, but rather it is about developing a robust capability for the objects that are most likely to require a short-warning-time, space-based response to provide critical information to decision makers. To support this work, Section 2 evaluates the potential asteroid threat population to determine the capabilities required to enable flyby reconnaissance for >90% of the population. Section 3 determines the measurements required during the flyby reconnaissance to support planetary defense decisions. Section 4 summarizes the driving requirements that result from these analyses and discusses the conclusions and next steps to advance this planetary defense NEO reconnaissance capability.

## 2. Assessment of the Potential Asteroid Threat Population

A synthetic threat population was created by adjusting the orbital phase of the population of the 2340 potentially hazardous asteroids existing at the time of the analysis [9] to achieve their minimum Earth distance in the early 2030s. Ballistic spacecraft trajectories were computed for each object in the synthetic threat population, and post-processing of the millions of possible trajectories was used to determine the range of flyby conditions to access the targets. Solar electric propulsion, gravity assists, and deep-space maneuvers were excluded in the analysis. Many of the trajectory parameters are correlated, and Figure 1 shows the combined results for two of the more challenging flyby conditions – approach solar phase angle and flyby speed.

To develop a robust planetary defense flyby reconnaissance capability, we assert that the capability should be successful for >90% of the potential NEO threat population. This assertion is motivated in part by the ≥90% completeness criteria for finding NEOs ≥140 m enshrined in the 2005 George E. Brown, Jr. Near-Earth Object Survey Act [10]. Figure 1 shows as dashed lines the flyby conditions that would allow a spacecraft to reach 80%, 90%, and 95% of the synthetic threat population. To achieve >90% coverage of the population, the flyby conditions include faster flyby speeds than experienced by many previously flown small body flyby encounters, which are also shown on Fig. 1.

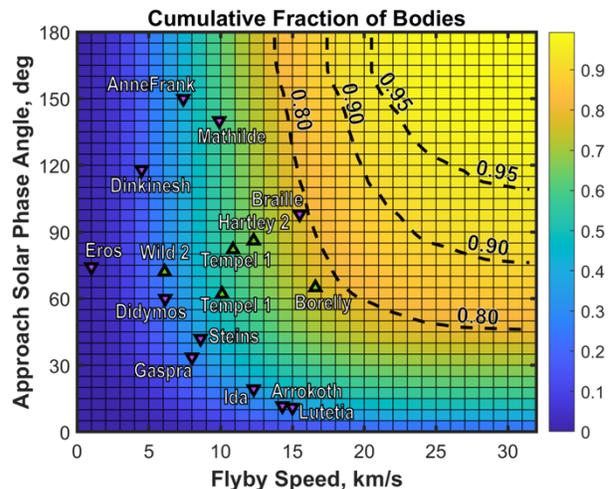

Fig. 1. Approach solar phase angle versus flyby speed for the synthetic threat population. The 0.9 line identifies flyby conditions that encompass 90% of the population., and the plot is colored by the cumulative fraction of bodies that can be reached within the combination of the flyby conditions. Previous asteroid (upside down pink triangles) and comet (green triangles) flybys are also plotted.

Figure 1 also shows that there is a trade-off in flyby speed and solar approach phase angle to be considered in selecting the flyby conditions that are required to be able to have successful reconnaissance of >90% of the population. A solar approach phase angle of 0° corresponds to the object being fully sunlit during approach, while a solar approach phase angle of 180° means the object is approached when not sunlit and in the dark. Ensuring the target can be identified by optical navigation to enable a successful flyby is a priority, and given the SMPAG and NITEP thresholds [1, 2], the object may be potentially as small as 50 m. We thus choose to limit the maximum approach solar phase angle to 90°, to ensure that at least half of the object is sunlit to support flyby navigation needs.

Table 1 gives our derived criteria for a flyby reconnaissance capability that encompasses >90% of the potential NEO threat population. Any individual flyby encounter is unlikely to reach all of the conditions listed in Table 1, but to have a capability that can be used for >90% of the potential NEO threat population, the flyby reconnaissance capability must be able to be successfully deployed for the conditions encompassed by Table 1. Note that this set of parameters is not the only set that meets 90% of the population. For example, one might decrease the time of flight at the expense of increasing the other parameters. This selection of parameters is challenging, but we view it as an achievable set that provides crucial reconnaissance information on a timeframe relevant for a potential Earth impacting object with a limited warning time.

Table 1. Flyby conditions for >90% completeness.

| Parameter | |
|---|---|
| Maximum time of flight | 2.5 yrs |
| Minimum solar distance | 0.9 AU* |
| Maximum solar distance | 2.0 AU* |
| Maximum approach solar phase angle | 90° |
| Maximum flyby speed | 25 km/s |
| Maximum launch C3 | 30 km$^2$s$^{-2}$ |

* AU = Astronomical Unit

It is worth noting that this analysis assumes a dedicated launch purposefully optimized for the needed flyby. Some previous studies [e.g., 11] have suggested that the option of storing a spacecraft in space and then deploying it once a potential Earth impacting object is discovered could be a strategy for planetary defense reconnaissance missions. However, our work shows that the store in space option with a single spacecraft is not viable to create a reconnaissance capability that can be successful for >90% of the potential NEO threat population. We did not evaluate multiple spacecraft options that are stored in space, though such options would clearly increase the cost and complexity. Even with a dedicated launch, the range of flyby conditions for which a reconnaissance mission must be designed to accommodate are demanding; removing the launch vehicle contribution to the thrust available to reach the potential impactor makes a large fraction of the potential NEO impact threat population unreachable with a single spacecraft solution that is stored in space.

## 3. Determination of Flyby Measurement Requirements

In addition to national and international policies, we reviewed all of the planetary defense exercises to date [12] to assess which information was most critical from a rapid reconnaissance mission. These exercises are carefully crafted to accurately represent the physics and timelines of a physically realizable hypothetical impactor.

### 3.1 Determine If the Object Will Impact the Earth

SMPAG and NITEP threshold criteria [1, 2] state that a reconnaissance mission should be conducted if the Earth impact probability rises to >1%. Thus, knowing where the object will hit the Earth, if at all, will be the top priority question for decision makers. Conducting a flyby will provide the needed orbit information to definitively determine whether the object will impact the Earth. The exact accuracy with which the impact location can be determined will be dependent on the specific NEO flyby conditions, but the potential impact location will likely be determined to within ~100 km following a flyby reconnaissance mission, as shown by the results from past planetary defense exercises [12].

### 3.2 Determine the Asteroid's Size

Size estimates for asteroids have large uncertainties when only ground-based optical telescopic data are available. Figure 2 demonstrates this using the estimated sizes for objects in planetary defense exercises conducted since 2013 [12]. Size here is defined as the diameter of a volume-equivalent sphere. Constraining the size for a small 50-m object is the most challenging case for this flyby reconnaissance and drives the associated requirements. Requiring ≥10 pixels across a 50-m object yields a pixel scale of ≤5 m, and a one-pixel uncertainty on each side of the measured extent yields a ±10 m 1σ uncertainty. Approach and departure imaging is needed to constrain the maximum size extent.

This pixel scale requirement assumes that the asteroid is roughly spherical; however, some near-Earth asteroids and comets are elongated. In order to be robust against such a scenario, approach and departure imaging needs to be complemented by imaging at a comparable pixel scale taken with a viewing geometry that differs by ≥60° and ≤120° from that of the approach viewing geometry. Refining the pixel scale requirement needed to account for the case of an elongated asteroid with a volume-equivalent diameter of 50 m is being further investigated. From Fig. 2 it is clear that this basic information on the size would vastly decrease the size uncertainties considered in the planetary defense exercise scenarios.

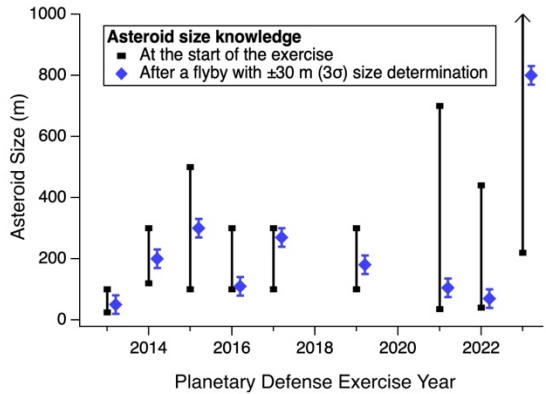

Fig. 2. Asteroid size knowledge from past planetary defense scenarios [12] are plotted along with the size determination obtained from a flyby reconnaissance mission.

Higher spatial resolution imaging, such as required to characterize the surface features and discussed in Section 3.5, would further improve the accuracy with which the NEO's size could be determined. This can be shown using the DART mission as an example: A study conducted prior to DART's impact event demonstrated that from DART's approach imaging alone, which was required to obtain images down to a resolution of 66 cm/pixel, the volume of roughly 150-m-diameter Dimorphos would be determined to better than 25% [13]. This volume error amounts to an 8% error in the asteroid's size. The final shape model of Dimorphos derived from DART's approach imaging that extended to 5.5 cm/pixel determined each axis of Dimorphos with an uncertainty of 4 m or better [14]. Flyby reconnaissance imaging obtained on both approach and departure, as well as imaging taken with a geometry ≥60° and ≤120° from that of the approach geometry, would provide multiple viewing geometries not available from DART's impact trajectory, further improving the ability to constrain the NEO's size.

It is also worth noting that a focus on resolution as a percentage of the size of the object (i.e., within 10%) does not fully capture the information that will be gained by this flyby capability. By deriving the requirements from the most stressing case, requirements will be exceeded for other cases. A 5-m spatial resolution will provide a smaller fractional size and volume uncertainty for objects > 50 m in diameter, with correspondingly smaller uncertainties for the estimated masses of these more destructive objects.

*3.3 Determine If the Asteroid Is Mostly Rocky or Metallic*

While direct measurements of the mass of a 50-m object during a flyby are beyond our current technologies, asteroid threat assessments are driven by having to consider the entire range of possible densities [12]. The asteroids with the most extreme densities have a Fe-Ni metal composition. Iron meteorites are only 5% of observed meteorite falls [15], but their high density makes them more massive for their size, which can result in substantially more damage. In the absence of direct mass measurements during a flyby reconnaissance encounter, acquiring measurements to determine whether the object is dominated by metal is required to provide crucial insight into the object's mass.

Determining whether an asteroid is mostly rocky or metallic has been done using Earth-based telescopic observations, both from radar [e.g. 16] and spectral [e.g. 17] observations. If a possible Earth impactor is identified, Earth- and space-based observatories can provide important compositional information. However, the viewing geometry and target brightness necessary to obtain the radar, visible/near-infrared spectroscopy, and/or mid-IR/thermal emissivity telescopic data that can reliably distinguish between rocky and metallic objects is not generally available for small objects, even from space telescopes. Thus, we cannot rely on having Earth-based observing opportunities to gain this key information, especially for scenarios which may involve 50-m sized objects with short-warning times.

Flyby reconnaissance can provide the data needed to distinguish mostly rocky from mostly metallic objects. Using visible/near-infrared spectroscopy, most rocky meteorites display characteristic silicate absorption bands [e.g. 18] while iron meteorite powders exhibit a featureless, red-sloped spectrum [19]. Such featureless spectra can also be found on asteroids that are organics-rich or have iron-poor silicates. Thus, visible/near-IR spectroscopy can be used to robustly identify non-metallic objects if bands are identified, but a featureless spectrum would leave some uncertainty about the rocky versus metallic nature of the object. Consequently, such visible/near-IR measurements can contribute to distinguishing between mostly rocky or metallic objects but are not sufficient measurements to rely on in isolation to gain this key information.

For thermal infrared spectral observations, dominantly rocky meteorites have considerably higher emissivity [20] than measured for iron powders [21], as shown in Fig. 3, making this a compelling approach to distinguish mostly rocky materials from those that are mostly metallic. Additionally, thermal inertia values can also distinguish between rocky and metallic meteorites, with metallic objects having higher thermal inertia values than rocky ones [22]. Thermal infrared measurements have been successfully demonstrated by previous asteroid reconnaissance missions, such as by the OSIRIS-REx mission at Bennu [23], whose data are also shown on Fig. 3. More recently, the Lucy Thermal Emission Spectrometer (L'TES) instrument obtained measurements during its flyby of the asteroid Dinkinesh [24], demonstrating the feasibility of thermal infrared spectral measurements during flyby reconnaissance.

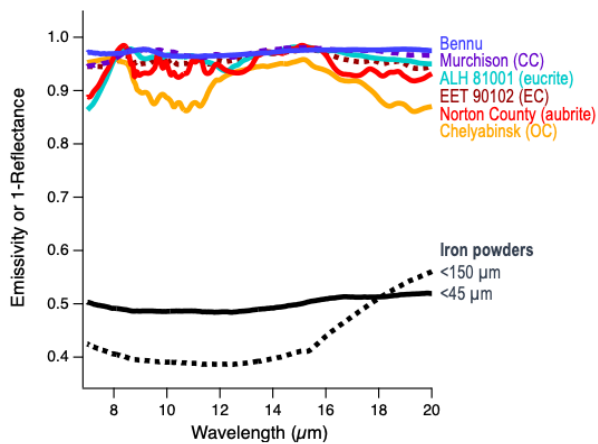

Fig. 3. Thermal infrared measurements for rocky meteorites [20] and for the rocky asteroid Bennu [23] are easily distinguished from measurements of metallic materials, such as iron powders [21].

Because of the importance of constraining the mass of any potential Earth impactor and the importance of density estimates to that mass constraint, a robust flyby reconnaissance capability should ideally include two independent measurements to confidently conclude whether the object is largely rocky or metallic. Thermal infrared spectral measurements have strong flight heritage and can compellingly discriminate between rocky and metallic bodies. Visible/near-infrared spectral measurements could be used to complement the thermal infrared measurements, supporting the thermal infrared conclusion by providing a spectrum with clear silicate bands or one that is featureless.

As an alternative secondary approach to constrain the rocky versus metallic nature of a NEO, measurements at sub-millimeter to radio wavelengths can probe the physical properties of the surface to a larger depth than shorter wavelength observations. The Microwave Instrument for Rosetta Orbiter (MIRO), a millimeter and submillimeter radiometer and spectrometer, obtained emissivity and thermal inertia measurements of the asteroids Steins [25] and Lutetia [26, 27] from flyby observations. The MIRO measurements provided insight into the nature of the asteroid surface properties that should be able to distinguish between a mostly rocky versus a metallic body, though the analysis done by the studies did not address this topic in particular, and supporting laboratory measurements and calibrations are needed.

Earth-based radar observations are an established means to identify metallic asteroids [16], and such measurements would be highly desirable to discriminate between a largely rocky or metallic body. While radar has not been carried on an asteroid mission to date, the concept of operations for obtaining spacecraft measurements analogous to Earth-based radar observations appears to be straightforward. In combination with imaging data, the radar cross-section for the target asteroid, and the radar albedo, can be determined. In a qualitative sense, the distance from which the target can be detected will be significantly larger for metal-rich targets compared to metal-poor ones. While studies have not been done, an instrument similar to the Miniature Radio Frequency (Mini-RF) on the Lunar Reconnaissance Orbiter [28] may be able to obtain the desired data. If it can be accommodated, the use of radar holds promise for a straightforward, secondary measurement of metal content.

Overall, a thermal infrared spectrometer, such as L'TES, is well suited to provide the data needed to discriminate between whether a NEO is mostly rocky or metallic during a flyby reconnaissance mission. More work is needed to trade off the best approach to also obtain a secondary, independent means to confirm this important distinction for planetary defense decision makers.

*3.4 Determine If There Is a Single Object or Multiple Objects of Concern*

Knowledge about whether there is a single object that may impact the Earth or multiple objects is fundamental information to inform a planetary defense mitigation approach. Observations of NEO binary asteroids show that the majority of secondaries have been found within distances of 5 primary diameters, and all are within 10 primary diameters [29]. Hence a reconnaissance mission would be required to search this region and identify any secondary objects of concern. The smallest object with associated mitigation actions in the NITEP protocols is 10 m is size, and thus, we adopt ≥10-m as the threshold for the secondary objects that a reconnaissance mission must discover as potential Earth impactors. Objects that are smaller than 10 m may also be of concern to some mitigation strategies, such as slow push mitigation techniques; however, any smaller objects could be identified in the early phases of such a follow-on mitigation mission. The priority for this initial reconnaissance flyby mission is to confidently determine if there is a single object or multiple objects that are on Earth-impacting paths. For an object's size to be determined, we require at least three pixels across the object, resulting in a required spatial resolution for the secondary search of ≤3 m/pixel.

The extent of the region that must be searched is 10 primary diameters, but the diameter of the asteroid is unknown and the capability developed must work for >90% of the potential impact threat population. For this requirement, the 50-m object is not the driving case but rather the largest object that may be needed to be encountered drives the required search region. Given that NEO Surveyor will discover >90% of NEOs ≥140-m [4], and the discovered objects will have longer warning times that could likely enable rendezvous reconnaissance missions, we use 140-m diameter objects to define the secondary search region. This approach yields a 1.4-km radius region around the primary asteroid that must be searched to detect any objects ≥10 m.

The required secondary search can be accomplished with a visible imager. This was nicely demonstrated during Lucy's recent flyby of the asteroid Dinkinesh [30], when the secondary Selam was discovered using the mission's imaging camera, as shown in Fig. 4. To confidently identify or rule out the presence of any secondary members ≥10 m, imaging during both the flyby's approach and departure phases to cover the potential secondary region at ≤3 m/pixel is required.

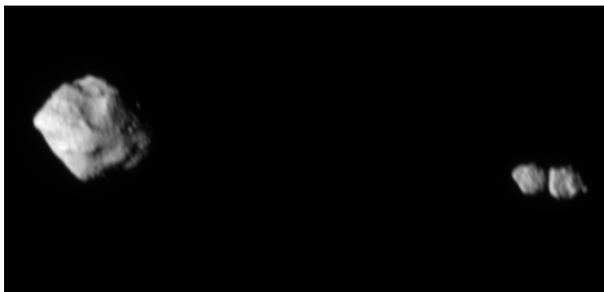

Fig. 4. Image acquired by the Lucy mission during its flyby of the asteroid Dinkinesh (720 m) that also revealed the secondary member Selam (with two lobes, with diameters of 210 m and 230 m), which orbits Dinkinesh at a distance of 3.1 km. The image was acquired 5.46 minutes after the spacecraft's closest approach of 431 km at a pixel scale of 7.56 m/pixel and a solar phase angle of 44.5° [30].

While all known secondaries for asteroids have been discovered within 10 primary diameters, a more conservative approach to ensure that there are no objects bound within the asteroid system is to search the object's Hill sphere, the region of gravitational influence of the primary asteroid. The largest asteroids of concern will be the driving case, with the Hill sphere depending upon mass. Using a 140-m diameter, metallic body, the Hill sphere radius extends to roughly 19 km. The Hill sphere was used to define the satellite search region for the OSIRIS-REx mission [31], though given the rendezvous rather than flyby nature of that encounter, such an extensive search was easily accommodated. For flyby reconnaissance with a speed of up to 25 km/s, more work is needed to investigate what fraction of the Hill sphere can be examined fully for secondary objects, with the motivation to cover the full Hill sphere region as the most robust option. While a larger search region is preferred, the required search time is longer and may detract from the limited time available to image the primary object.

*3.5 Determine the Asteroid's Surface Characteristics*

Determining the surface characteristics of the NEO target is important to inform potential mitigation options. DART's approach imaging of 150-m Dimorphos [32], with images acquired every second with the same viewing conditions but increasingly higher spatial resolution, provides an appropriate dataset for investigating the surface imaging resolution requirement for flyby reconnaissance of a small NEO. Figure 5 shows three images acquired by DART, cropped to show the same 50-m-diameter surface region but at different pixel scales. Actual imaging of a 50-m diameter sunlit body would show lighting variations across the scene, unlike the cropped images in Fig. 5. However, the DART images provide a means to evaluate surface features that can be identified with imaging at different pixel scale resolutions. Figure 5 shows that image resolutions of ≤0.5 m/pixel are needed to confidently identify boulders on Dimorphos and hence to characterize the nature of the surface.

Characterizing features on the asteroid's surface at the scale of a potential mitigation technique can inform the mitigation approach. The DART kinetic impactor spacecraft was roughly 2 m in size [32], and thus, imaging a NEO surface at 0.5 m/pixel would enable confident identification of surface features at the size of a potential mitigation spacecraft. Approach and departure imaging is required at this resolution to provide insight into the surface characteristics of the object for the sunlit surface areas.

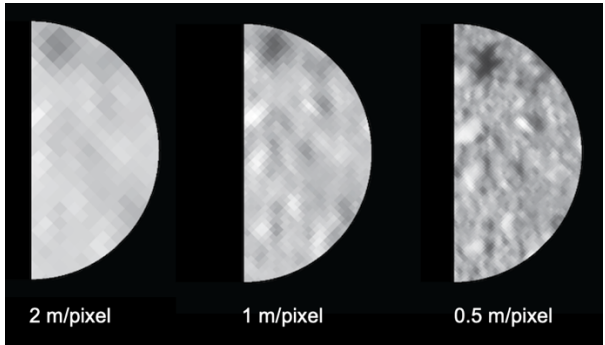

Fig. 5. Three images acquired by the DART mission during its approach to Dimorphos [32], with each image showing the same area of the asteroid's surface that is 50 m in the vertical dimension but at different pixel scales.

## 4. Results and Conclusions

From our analysis of the potential NEO threat population and flyby measurement requirements, we define four major requirements for a planetary defense flyby reconnaissance demonstration mission:

1. Enable a flyby of >90% of the potential asteroid threat population.

As noted repeatedly, in planetary defense, you don't pick the asteroid — the asteroid picks you. This reality drives the development of the required capability for a reconnaissance flyby mission. Our results show that to enable a flyby of >90% of the potential asteroid threat population, flyby speeds of up to 25 km/s and approach solar phase angles as high as 90° are required to be accommodated.

2. Demonstrate the flyby reconnaissance for a ~50 m NEO.

A ~50 m diameter object is the most likely threat that the Earth will face that will warrant a space-based reconnaissance mission by SMPAG [1] and NITEP [2] thresholds. A demonstration mission is essential for validating that the as-designed system can reliably perform when needed. If an asteroid threat to the Earth was identified, the flyby encounter would represent a critical, singular opportunity to gather the reconnaissance information required for decision makers to develop an effective mitigation plan. A preventable spacecraft failure would be intolerable. The demonstration mission must identify any potential failures and test the limits of the spacecraft's performance in a stressing flyby. A stressing flyby case would involve a ~50-m (absolute magnitude ~26) object with a challenging fast flyby speed of near 25 km/s. The demonstrated flyby reconnaissance capability could then also be applied to the less challenging cases of a larger object or slower flyby speed, should such a need arise in the future.

3. Obtain the information needed to determine if and where the object would impact the Earth.

The SMPAG [1] and NITEP [2] criteria state that a spacecraft reconnaissance mission should be considered if the Earth impact probability is >1%. Determining if the object will impact the Earth and if so, constraining the impact swath location on the planet, is a top priority, and hence warrants its own top-level requirement. If it is determined that the object will not impact the Earth, determining the NEO's properties, as defined in the next requirement #4, becomes much less important.

4. Determine key properties of the asteroid to inform decision makers.

While in any threat situation, as much information as possible is always desired, our analysis identifies four key properties that are the priorities, with discussion of their implementation and measurement requirements.

4a. *Determine the asteroid's size*: Imaging on approach and departure, complemented by imaging with a viewing geometry that differs by ≥60° and ≤120° from that of the approach, at ≤5 m/pixel is required. Imagers flown on multiple asteroid missions, including on DART [32] and Lucy [30], have demonstrated this capability successfully at lower flyby speeds.

4b. *Determine if the asteroid is mostly rocky or metallic*: Obtaining thermal infrared spectral data is a flight-proven technique, such as on OSIRIS-REx [23] and Lucy [24], that can discriminate between mostly rocky and metallic objects. Because of the importance of making this distinction to inform planetary defense mitigation techniques, two independent measurement approaches should ideally be used to make this determination during the flyby reconnaissance. For a second means to make this distinction, visible/near-infrared spectral measurements could be used to complement the thermal infrared measurements but may not yield diagnostic results on their own. Alternatively, a millimeter and submillimeter radiometer and spectrometer, such as MIRO flown on Rosetta [25–27], can obtain emissivity and thermal inertia measurements of the asteroid, which provides insight into the surface properties, but has not been shown to specifically distinguish mostly rocky from mostly metallic objects. Earth-based radar observations are a proven technique to identify metallic asteroids [15]. Collection of such radar measurements has not yet been demonstrated for a small object encountered by a fast flyby, but utilizing an approach such as Mini-RF flown on the Lunar Reconnaissance Orbiter [28] is worthwhile to investigate.

4c. *Determine if there is a single object or multiple objects of concern*: Imaging on approach and departure at ≤3 m/pixel is required, covering the region that extends to 1.4 km from the primary asteroid. As with #4a, imagers flown on multiple asteroid missions have demonstrated this general capability but at lower flyby speeds. A more conservative approach to ensure that there are no objects bound within the asteroid system is to expand the search region to cover the full Hill sphere, which would mean covering the region within roughly 19 km of the asteroid at ≤3 m/pixel.

4d. *Determine the asteroid's surface characteristics*: Imaging on approach and departure at ≤0.5 m/pixel is required to identify surface features on the scale of a potential mitigation spacecraft. DART acquired images at this pixel scale and better during its approach to Dimorphos [32], though the flyby speed was lower and there were no departure imaging opportunities.

These four overarching requirements define a planetary defense flyby reconnaissance demonstration mission that is not about just flying by an asteroid but rather is about developing a robust capability for the objects that are most likely to require a short-warning-time, space-based response to provide critical information to decision makers. The next steps are to take these requirements driven by planetary defense needs and to investigate implementation options for a flyby demonstration mission. In particular, we anticipate that navigation may be one of the largest technical challenges for this concept, given the fast flyby speed, high approach solar phase angle, and small, potentially low-albedo object. However, we do not have the luxury of choosing the asteroid, so addressing this challenge is necessary to advance our planetary defense preparedness capabilities.

**Acknowledgements**

This work was funded internally by the Johns Hopkins University Applied Physics Laboratory.